# Three-dimensional quantum Hall effect and metal-insulator transition in ZrTe$_5$

Fangdong Tang[1,8]*, Yafei Ren[2,3]*, Peipei Wang[1], Ruidan Zhong[4], J. Schneeloch[4], Shengyuan A. Yang[5¶], Kun Yang[6], Patrick A. Lee[7], Genda Gu[4], Zhenhua Qiao[2,3#], Liyuan Zhang[1§]

1 Department of Physics, Southern University of Science and Technology, and Shenzhen Institute for Quantum Science and Engineering, Shenzhen, 518055, China

2 ICQD, Hefei National Laboratory for Physical Sciences at Microscale, and Synergetic Innovation Center of Quantum Information and Quantum Physics, University of Science and Technology of China, Hefei, Anhui 230026, China

3 CAS Key Laboratory of Strongly-Coupled Quantum Matter Physics, and Department of Physics, University of Science and Technology of China, Hefei, Anhui 230026, China

4 Condensed Matter Physics and Materials Science Department, Brookhaven National Laboratory, Upton, New York 11973, USA

5 Research Laboratory for Quantum Materials, Singapore University of Technology and Design, Singapore 487372, Singapore

6 National High Magnetic Field Laboratory, Florida State University, Tallahassee, Florida 32306-4005, USA

7 Department of Physics, Massachusetts Institute of Technology, Cambridge, Massachusetts 02139, USA

8 Department of Physics, Renmin University of China, Beijing 100872, China

*These authors contributed equally to this work. Correspondence and requests for materials should be addressed to S.A.Y. (email: shengyuan_yang@sutd.edu.sg), Z.Q. (email: qiao@ustc.edu.cn), and L.Z. (email: zhangly@sustc.edu.cn).

**Symmetry, dimensionality, and interaction are crucial ingredients for phase transitions and quantum states of matter. As a prominent example, the integer quantum Hall effect (QHE) represents a topological phase generally regarded as characteristic for two-dimensional (2D) electronic systems, and its many aspects can be understood without invoking electron-electron interaction. The intriguing possibility of generalizing QHE to three-dimensional (3D) systems was proposed decades ago, yet it remains elusive experimentally. Here, we report for the first time clear experimental evidence for the 3D QHE, observed in bulk $ZrTe_5$ crystals. Owing to the extremely high sample quality, the extreme quantum limit with only the lowest Landau level occupied can be achieved by an applied magnetic field as low as 1.5 T. Remarkably, in this regime, we observe a dissipationless longitudinal resistivity $\rho_{xx} \cong 0$ accompanied with a well-developed Hall resistivity plateau $\rho_{xy} = (1 \pm 0.1)\frac{h}{e^2}\left(\frac{\lambda_{F,z}}{2}\right)$, where $\lambda_{F,z}$ is the Fermi wavelength along the field direction (z axis). This striking result strongly suggests a Fermi surface instability driven by the enhanced interaction effects in the extreme quantum limit. In addition, with further increasing magnetic field, both $\rho_{xx}$ and $\rho_{xy}$ increase dramatically and display an interesting metal-insulator transition, representing another magnetic field driven quantum phase transition. Our findings not only unambiguously reveal a novel quantum state of matter resulting from an intricate interplay among dimensionality, interaction, and symmetry breaking, but also provide a promising platform for further exploration of more exotic quantum phases and transitions in 3D systems.**

Since its discovery in 1980, the QHE has been established and well understood in a variety of 2D electron systems, including the traditional 2D electron gas[1,2], and 2D materials like graphene[3,4], etc. The hallmark of QHE is that the Hall conductivity $\sigma_{xy}$ takes precisely quantized values as $ve^2/h$ while the longitudinal conductivity $\sigma_{xx}$ vanishes[1,2]. Here, the prefactor $v$ is the filling factor which counts the number of filled Landau levels, $e$ is the elementary charge, and $h$ is Plank's constant. Soon after its discovery, the possibility to extend the QHE to 3D systems was speculated[5-8]. The most straightforward strategy is to stack multiple copies of 2D QHE layers along the z axis. With negligible interlayer coupling, a quantized 3D Hall conductivity $\nu \frac{e^2}{h}\frac{1}{l}$ would be expected, assuming $l$ is the 2D QHE layers spacing separation. This idea was indeed realized in semiconductor superlattice systems[9,10]. However, such weakly coupled layer systems are still of 2D nature (often referred to as quasi-2D), whose topology of Fermi Surface (FS) is opening, and qualitatively different from a 3D enclosed FS. This distinct topology of FS has clear experimental indications for quasi-2D and (ansiotropic) 3D systems. For example, in the presence of a magnetic field B, the electron's cyclotron motions for 3D FS are always in closed semi-classical orbits, regardless of B's direction. Whereas, for an opening quasi-2D FS, the electrons cannot complete a cyclotron motion where the orbits are open in certain B directions.

To distinguish a genuine 3D QHE from such quasi-2D QHE, one needs to compare two energy scales[11]: the bandwidth $W_z$ along the magnetic field (z) direction and the Fermi energy $E_F$. If $W_z < E_F$, the system is quasi-2D. On the other hand, if $W_z \geq E_F$, then the system is 3D.

If we increase the interlayer coupling to make the layered structure a "genuine 3D system", the bulk will often become conductive due to dispersion along z direction, and the quantization of $\sigma_{xy}$ will be lost. Nevertheless, in a pioneering theoretical work[5], Halperin showed that for a general 3D system with a periodic potential, as long as the Fermi level lies in an energy gap, the Hall conductivity will take the form of (assuming the magnetic field is in the z direction)

$$\sigma_{xy} = \frac{e^2}{2\pi h} G_z \qquad (1)$$

where $G_z$ is the z-component of a reciprocal lattice vector for the potential, with $\sigma_{xx} = 0$ in the meantime. Recall that the magnetic field quantizes the electron motion in the $xy$-plane into confined cyclotron orbits, but does not affect its motion in the field (z) direction. This leads to continuous 1D Landau bands which disperse with $k_z$. For realistic 3D crystalline materials, if the energy gap stated in the condition for Eq. (1) comes from the gap between the Landau bands (in which case $G_z$ corresponds to the crystal lattice periodicity), the required magnetic field $B$ will be at least thousands of Tesla, which is unattainably large.

There is another possibility. The energy gap may be opened by interaction effects that drive the Fermi surface instability, such as the formation of charge density wave (CDW)[12,13,14], spin density wave (SDW)[15], or excitonic insulator states[16]. In such cases, $G_z$ corresponds to the period of the formed "superstructure" which generally differs from the lattice periodicity. This possibility of an interaction-driven 3D QHE is further facilitated by the magnetic field, which suppresses the electrons' kinetic energy in the $x$-$y$ plane, making them effectively "quasi-1D" particles. The reduced dimensionality will enhance the interaction effects. Particularly, in the so-called extreme quantum limit with only the lowest Landau band (partially) occupied, we would have the celebrated Peierls problem for a single partially-filled quasi-1D band, which possesses the well-known Fermi surface instability with arbitrarily small interaction strength, due to its perfect Fermi surface nesting.

This intriguing possibility of 3D QHE has been actively pursued in crystalline materials in the past, for example, inorganic Bechgaard salts[17,18], $\eta$-$Mo_4O_{11}$[19], $n$-doped $Bi_2Se_3$[20], $EuMnBi_2$[21] and $SrTiO_3$[22]. Although traces of the emergence of a gap were observed in these experiments, the crucial evidence for the QHE, i.e., the vanishing $\rho_{xx}$, had not been observed. In ~~fact~~ addition, those reported experiments are lack of some clear in-plane quantum oscillations, which are essential evidence for 3D nature.~~.~~ Thus, a definitive testimony for the 3D QHE has not yet been demostrated.

The main challenges of observing 3D QHE are the following. First, the condition for achieving the extreme quantum limit requires ultralow carrier density or extremely high magnetic field, which is very hard to achieve for most realistic materials. Second, the opening of a sizable interaction gap (compared with the measurement temperature) requires the sample to have very high quality. Otherwise, the formed CDW or SDW ordering would be easily suppressed by impurities or defects.

Zirconium pentatelluride ($ZrTe_5$) offers a desirable platform for investigating 3D QHE, as we demonstrate below. $ZrTe_5$ has an orthorhombic layered structure with space group *Cmcm* (No. 63), as schematically shown in Figure 1**a**. Here, the three principal crystal axes $a$, $c$, and $b$ are labeled as corresponding to the directions ($x$, $y$, $z$). The material was studied in the past for its unusual thermoelectric properties[23], and recently it attracts revived interest due to its possible nontrivial band topology [24,25,26]: Its conduction and valence bands nearly touch, forming a Dirac-point-like band structure around the Γ point[27-30]. The material enjoys the following advantages for our purpose. It can be made very pure, with negligible intrinsic defects, which enables the achievement of ultralow carrier density and ultrahigh mobility. As we shall show, the extreme quantum limit can be obtained by a small $B$ field as low as 1.5 T. The Dirac like spectrum helps to increase the mobility and produce the unconventional zeroth Landau band. In addition, the strong structural anisotropy gives rise to a relatively large effective mass along $z$, which further enhances the interaction effects when a magnetic field is applied along this direction.

Our experiment has been performed on four different bulk ZrTe$_5$ samples. Hereafter we shall present the explicit results for one particular sample S#2, and the main features discussed below are also manifested by other samples. The inset of Figure 1**b** illustrates the setup of our transport measurement. It takes the standard Hall bar geometry, with current $I$ running along the $x$ direction through a 3D bulk ZrTe$_5$ sample. We first characterize the transport property in a wide temperature range from $T = 0.6$ K to 200 K. The measured temperature dependence of the resistivity $\rho_{xx}(T) = \frac{W \cdot d}{L} R_{xx}(T)$ in zero magnetic field is shown in Figure 1**b**. Here, $W$, $L$, and $d$ respectively denote the width, length, and thickness of the sample defined in the longitudinal and Hall transport measurement. One clearly observes an anomalous peak around temperature $T_p \sim 95$ K, which is attributed to a transition of carrier type. Furthermore, this is verified by the standard low magnetic field Hall measurements in Figure 1**c**, from which we extract the dominated carrier's density ($n = \frac{dB}{|e|d\rho_{xy}}$) and mobility $\mu = \frac{1}{|e|\rho_{xx} \cdot n}$ as functions of the temperature. The results are plotted in Figure 1**d**. One observes that the mobility is extremely high at low temperature, about $10^5$ cm$^2$V$^{-1}$s$^{-1}$ at 0.6 K. It decreases rapidly as temperature increases, and notably the carrier density changes sign around $T_p$, indicating that the system transforms from an "electron" metal below $T_p$ (with

d$\rho_{xx}$/d$T > 0$) to a "hole" insulator above $T_p$ (with d$\rho_{xx}$/d$T < 0$). This observation is in good agreement with the previous ARPES measurements[31], which had been explained as a temperature-driven Lifshitz transition around $T_p$. In the following, we will focus on the low-temperature regime where mobility is high and transport is dominated by electron carriers, and most results are obtained at 1.5 K unless stated otherwise.

The previous experiments and first-principles calculations have indicated that for the lightly *n*-doped case, bulk ZrTe$_5$ has a single electron pocket around the Γ point. This picture is verified in our experiment (See Supplementary Information). The morphology and topology of the Fermi pocket are probed by the Shubnikov–de Haas (SdH) oscillations in the magnetoresistance. We rotate the direction of the applied *B* field with respect to the crystal axes, and measure SdH oscillations in a series of rotation angles. Figure 2 **a** shows the results for *B* field along the three principal directions *x*, *y*, and *z*. One observes that the oscillations start at very small fields, indicating the high mobility of the sample. For example, for *B* field along the *z* axis, the oscillation can be discerned at $B_{int} \sim 0.067$ T, and the Hall mobility can be roughly estimated as $1/B_{int} \sim 150{,}000$ cm$^2$V$^{-1}$s$^{-1}$, which agrees with the result from Hall measurement, and is about three times higher than the previous reports [32,33].

The frequency of the SdH oscillation ($B_{F,i}$) is determined by the extremal cross-sectional area ($S_{F,i}$) of the Fermi surface normal to the field direction via the Onsager realtion: $B_{F,i} = S_{F,i}\left(\frac{\hbar}{2\pi e}\right)$, where *i* denotes the field direction. In our measurement, we find only a single dominant frequency for each field direction, consistent with the picture of a single electron pocket with a regular convex shape, as we mentioned before. The extracted oscillation frquencies for *B* field along the three pricipal directions are $B_{F,x} = 15.7 \pm 0.2$ T, $B_{F,y} = 9.2 \pm 0.1$ T, and $B_{F,z} = 1.18 \pm 0.02$ T, respectively. As shown in Figure 2 **e**, we are assuming that electron pocket is of ellipsoidal shape, which is reasonable for orthorhombic crystal symmetry and at low carrier density, we have $S_{F,z} = \pi k_{F,x} k_{F,y}$, where $k_{F,i}$ is the Fermi wave-vector along the *i* direction, and similar relations hold for $S_{F,x}$ and $S_{F,y}$. Thus, from the SdH oscillation, one can readily extract the three Fermi wave-vectors $k_{F,i}$ ($i = x, y, z$). Their values are listed in Table *I*. The result shows that the pocket is quite small, indicating the ultra-low carrier density. And $k_{F,z}$, which is along the stacking direction, is an order of magnitude larger than $k_{F,x}$ and $k_{F,y}$, reflecting the strong structural anisotropy. The associated effective mass along $z$ is ~ 2.5 $m_e$, about two to three orders of magnitude larger than the other two directions, which tends to enhance the interaction effects under a strong *B* field.

Since $k_{F,z}$ is a crucial factor in the following discussion, we also estimate its value using an alternative approach. The bulk carrier density has been determined by the low-field Hall measurement, and in our case, it is related to the Fermi wave-vectors via $n_{3D} = k_{F,x} k_{F,y} k_{F,z}/(3\pi^2)$. Hence, $k_{F,z}$ can be obtained as $k_{F,z} = 3\pi^2 n_{3D} \hbar/(2eB_{F,z})$. For instance, for sample S#2, $k_{F,z}$ estimated using this approach gives a value of ~ $0.056 \pm 0.001$ Å$^{-1}$, which agrees well with the value ~ $0.061 \pm 0.004$ Å$^{-1}$ obtained solely from the SdH measurements. This in turn supports the assumption that the pocket has ellipsoidal shape.

We also mention that the phase offset of the SdH oscillation contains information about the Berry phase for the electron orbit. In Figure 2 **b**, via the standard Landau-level fan diagram analysis, we obtained the SdH oscillation frequency $B_F$ Vs. the tilted angles $\alpha$ and $\beta$, as shown in Figure 2 **c** and **d**. A 2D planar FS is fitted by the red curve assuming $B_F^{2D} = B_{Fz}/\cos\theta$, and blue curve is for a 3D ellipsoidal FS following $B_F^{3D} = B_{Fz} B_{Fi}/\sqrt{(B_{Fz}\sin\theta)^2 + (B_{Fi}\cos\theta)^2}$, with $\theta = \alpha$ or $\beta$, and $i = x$ or $y$. It is deviated as $\theta > 75°$ for 2D planar fitting, but fitting perfectly with a 3D ellipsoidal closed FS in whole range.

Moreover, we find that the Berry phases for *B* field along the $x$, $y$, and $z$ directions are respectively 0, 0, and $\pi$. Comparison with theoretical calculations (provided in Supplementary Information) indicates that the bulk ZrTe$_5$ sample in our experiment is likely to be in a (doped) weak topological insulator phase. It should be noted that for the *n*-type ZrTe$_5$ samples studied here, we find that the observed transport features are dominated by the 3D bulk carriers, whereas possible impacts from the surface states are negligible.

Having ~~determined~~ confirmed the topology of 3D Fermi surface, we then focus on the intermediate field range with $B \sim 0 - 12$ T along the $z$ direction. The zoom-in plot for $\rho_{xx}$ and $\rho_{xy}$ versus $B$ field strength in this field range is shown in Figure 3**a** for sample S#2. Here, the oscillation in $\rho_{xx}$ is labeled by the filling factor $\nu$ (= 1, 2, 3, 4 ...) with values extracted from the Landau-level fan diagram. (This periodic in $1/B$ oscillation can also be clearly observed in the plot of $d\rho_{xy}/dB$ vs. $1/B$, as shown in Figure 3 **f** ). There are several key observations (which also exist for the other samples). First, the extreme quantum limit that goes below $\nu = 1$ is achieved in ZrTe$_5$ at a very small field $\sim$ 1.3 T, much lower than most quantum Hall systems studied to date. Second, like in QHE, there appears clear correlation between $\rho_{xx}$ and $\rho_{xy}$: the dips in $\rho_{xx}$ correspond to relatively flat plateaus in $\rho_{xy}$, and the peaks in $\rho_{xx}$ correspond to the transition regions between the plateaus in $\rho_{xy}$. Last and the most remarkable feature is that after entering the extreme quantum limit, the last dip in $\rho_{xx}$ becomes vanishingly small: $\rho_{xx} \cong 0$ in a range of $B$ field around 2 T , accompanying a well-developed plateau in $\rho_{xy}$, and $\rho_{xy} \gg \rho_{xx}$. This is exactly the hallmark of the 3D QHE predicted by Halperin.

It must be pointed out that for the QHE observed here, the system bandwidth $W_z$ ($\sim$400 meV) [24] is much larger than the Landau level spacing $\Delta_B$ ($\sim$25 meV at 2 T). Hence, the QHE observed in our experiment is indeed of 3D nature, distinct from those quasi-2D systems consisting of weakly coupled 2D QHE layers.

To understand the origin of $\rho_{xy}$ planteaus, we first note that the value of the Hall resistance $R_{xy} = \frac{\rho_{xy}}{d}$ at the quantized plateau (e.g. R$_{xy} \cong$ 1.35 Ω, a thickness of d $\cong$ 110 μm for the sample S#2) is much less than the Klitzing constant $\frac{h}{e^2} \approx$ 25.812 kΩ by a factor of about $10^{-4}$, indicating that the effect originates from the bulk, rather than the 2D top and bottom surfaces. In Halperin's theory, the suppressed $\rho_{xx}$ ($\sigma_{xx}$) corresponds to an energy gap opened in the spectrum, in which the $\rho_{xy}$ ($\sigma_{xy}$) takes quantized values. From Eq. (1), the corresponding expression for the Hall resistivity is $\rho_{xy} = \frac{h}{e^2} \lambda_Q$, where $\lambda_Q$ is the period for the periodic potential in real space (possibly divided by an integer number), and can be extracted from the value of the measured quantized plateau in $\rho_{xy}$. We find that the value of $\lambda_Q$ varies from sample to sample. For sample S#2, $\lambda_Q \approx 5.9$ nm. Evidently, the period $\lambda_Q$ is much larger than the lattice constant along z (~ 1.45 nm), and the energy gap cannot be explained as arising from the lattice potential.

A key observation is made by noticing that the value of $\lambda_Q$ is close to half of $\lambda_{F,z}$, i.e. the Fermi wavelength along z. Surprisingly, this observation applies to *all* the samples that we studied. In Figure 3 **b**, we plot the values of $\lambda_Q$ and $\lambda_{F,z}$ for the four samples we have studied in Table *II*. One can see that, although $\lambda_Q$ and $\lambda_{F,z}$ vary from sample to sample (mainly due to the variation in carrier density), all the data points are sitting close to the straight line representing the relation $\lambda_Q = \lambda_{F,z}/2$. Since the wavelength $\lambda_{F,z}/2$ corresponds to the wave-vector of $2k_{F,z}$, namely, the span of the Fermi surface along z, this strong correlation between $\lambda_Q$ and half of $\lambda_{F,z}$ clearly points to a Fermi surface instability. Indeed, it has been long believed that a strong magnetic field can drive various Fermi surface instabilities for a 3D electron gas. The essential physics is the following. The applied *B* field quantizes the electron motion in the plane perpendicular to the field, effectively suppressing the in-plane kinetic energy, whereas the motion along the field is not affected, thus making the system behave as a quasi-1D system. The original 3D band structure is turned into the 1D Landau band spectrum with only $k_z$ dispersion, as illustrated in Figure 3 **c**. With sufficiently strong field (here about 1.3 T), the system enters the extreme quantum limit, where only the lowest ($N = 0$) Landau level is occupied, and the Fermi surface becomes only two points with perfect Fermi surface nesting. It is well known that such quasi-1D systems possess pronounced Fermi surface instabilities driven by interaction effects towards various insulating phases, such as CDW, SDW, excitonic insulators, etc. For the current system, there is only a single electron Fermi pocket, and the lowest Landau level has no spin degeneracy[34] (owing to both Zeeman splitting and strong spin-orbit coupling), hence SDW and excitonic insulator states are unlikely. The only reasonable candidate is therefore a CDW state. In such a CDW state, one expects periodic electron density modulation along the z direction, with a period corresponding to the nesting wave-vector close to $2k_{F,z}$ right after entering the extreme quantum limit, which is in good agreement with the experimental observation.

Our result, therefore, indicates an interaction-driven 3D QHE. Here, the interaction effects are enhanced by several factors. One is the Landau quantization by the magnetic field, which effectively reduces the dimensionality of the electronic system. The second is the structural anisotropy, resulting in relatively small dispersion along the $z$ axis. And the third one is the ease in achieving the extreme quantum limit due to the low carrier density and single electron pocket, where perfect Fermi surface nesting can be realized. We noticed that some previous experiments on ZrTe$_5$ had multi sequences of quantum oscillations in $\rho_{xx}$ (also $\rho_{xx} \gg \rho_{xy}$)[27,28], which is possibly due to the different sample qualities arising from different sample growth methods.

Moreover, we note that $\rho_{xx}$ approaches zero only in a small interval of field strength, e.g. 1.7 T to 2.2 T for Sample S#2, and starts to increase with further increasing $B$ field. This indicates that the energy gap induced by the CDW may cross the Fermi level only over a limited range of magnetic field. The wave-vector $k_{F,z}$ in the extreme quantum limit should depend on $B$, since the Landau level degeneracy is proportional to $B$. In the simplest picture, then it seems that the period of CDW $\lambda_Q$ and hence $\rho_{xy}$ would be field dependent, apparently contradicting the relatively flat plateau that is observed. These observations can be understood by noticing that the values of $\lambda_Q$ are close to integer multiples of the interlayer spacing along $z$ (recall that there are two layers per unit cell, and the interlayer distance is about 0.725Å): for S#1, S#2, S#3, and S#4, these values are within error equal to 4, 8, 7, and 7 times, respectively (see Table *II*). Given the layered structure in the $z$ direction, it is reasonable that such commensurate CDWs are more stable than the incommensurate ones[35,36], and once a commensurate CDW is formed, it is likely to be pinned by the interaction with lattice for a range of $B$ field strength. Further evidence for the field-induced CDW state comes from nonlinear transport[37,38] along $z$, for which one expects a non-Ohmic behavior arising from a sliding CDW, when the applied bias voltage or current reaches the depinning threshold. This has indeed been observed in our experiment (see Supplementary Information Sec. 7).

In Figure 3 **d**, we plot the angular dependence of Hall resistivity $\rho_{xy}(B)$. With rotating $\beta = 0°, 15°, 30°, 60°$, the tilting angle between $B$ and $y$ axis, we find the quantization values of $\rho_{xy}(B)$ are consistent with $\sim \frac{h}{e^2}\frac{\lambda_{F,z}}{2}$. Similar results were obtained in $\alpha < 70°$ (the tilting angle between $B$ and $x$ axis, see Supplementary Information), where $\rho_{xy}(B)$ depends only on the perpendicular component of the field $B_\perp = B\cos\beta$ (or $\alpha$). Figure 3 **e** shows $\rho_{xy}(B)$'s temperature dependence ($T = 1.5, 3, 5, 7, 9, 12, 15$ K) with applied $B//z$ direction, and the plateaus of $\rho_{xy}(B)$ survive even at $T = 15$ K. Using standard methods to determine the quantization's features, we plot the $d\rho_{xy}/dB$ (or $-d^2\rho_{xx}/dB^2$, see Supplementary Information) as a function of $1/B_\perp$ shown in Figure 3(f). The pronounced peaks are periodic in $1/B_\perp$, in all angular and temperature dependence data from Figure 3 **d** and **e**, which is consistent with the results of $-d^2\rho_{xx}/dB^2$. Then, we find that the positions of dips correspond to the filling factor $\nu = 1 \sim 5$. With this identification, we note an apparent peak corresponding to the fractional filling factor $\nu = 1/3$ in the quantum limit region for all 4 samples. The correspondence of this peak with a fraction of Landau index is striking, and this signature provides a possible precursor for exploring the Laughlin type of 3D fractional quantum Hall state in ZrTe$_5$ materials.

Next, we turn to explore the temperature dependence of $\rho_{xx}(B)$ at even higher field strength. Figure 4 **a** shows the transport measurement up to 13 T with temperature $T$ varying from 1.5 K to 15 K. One observes that $\rho_{xx}(B)$ stops to oscillate for $B > 3$ T, and it increases rapidly without any sign of saturation, achieving a colossal magnetoresistance more than 15000% at 13 T. Strikingly, when plotting the $\rho_{xx}(B)$ curves measured at different temperatures (1.5 K to 20 K) together, one finds that all the curves share a common crossing point at a critical magnetic field $B_c = 6.71$ T (inset is the zoom-in image of Figure 4 **a** around the critical point). We also plotted the $\rho_{xx}$ as a function of $T$ with $B$ varying from 2 to 13 T in figure 4 **b**. The $\rho_{xx}$ is monotonically increasing as $T$ dropping in the region above $B_c$, expected for an insulator. This feature also appears in Samples S#1 and S#3, and it clearly signals a critical point for a metal-insulator transition: the system is metallic below $B_c$ and is insulating above $B_c$.

Typically, for $B$-field induced metal-insulator transitions, the isotherms of $\rho_{xx}$ have a universal scaling with the parameter $(B - B_c)T^{-1/\zeta}$. In Figure 4 **c**, we perform such scaling analysis, and indeed, all the isotherms fall onto a single curve as a function of $|B - B_c| T^{-1/\zeta}$, with a fitted critical exponent $\zeta \approx 5.5$.

We are unable to determine the nature of the insulating state above $B_c$ at present. There are at least two possibilities. The first candidate is the formation of Wigner crystals. We note that the values of $B_c$ corresponding to the critical filling factor $v_c \sim$ 0.182, 0.173, and 0.193, for Sample S#1, S#2, and S#3, respectively, which are all below 1/5. This is consistent with the common belief that Wigner crystals would form below the critical filling factor of 1/5 [39,40,41]. The second possibility is the localization due to impurities and defects [42-45]. For high-quality samples, the main source of defects are the dopants. Here Landau orbit size gets progressively smaller in a strong magnetic field, leading to carrier localization on these defects.

Finally, we summarize our findings with a phase diagram in the $B$-$T$ plane shown in Figure 4 **d**. First, from Figure 1**c**, we identify a "hole" insulating state at $T > 95$ K, which converts to "electron" metallic states at $T < 95$ K. In the presence of magnetic field $B$, Landau Level quantization leads to CDW or other Fermi surface instabilities in region *I* at low temperatures; this region may actually host a sequence of 3D integer QHE which we have evidence for, but are unable to establish due to the finite $\rho_{xx}$. In the extreme quantum limit region *II*, we have clear evidence of a 3D QHE. We have seen suggestive signature for a 3D 1/3 fractional quantum Hall state in region *III*. At last, we find a metal-insulator transition with further increasing $B$. We note this is a zero temperature (quantum) phase transition, which needs to be explored further in the future. The finite $T$ boundaries separating metallic and insulating phases, on the other hand, represent crossovers.

## References

1   Vonklitzing, K., Dorda, G. & Pepper, M. New Method for High-Accuracy Determination of the Fine-Structure Constant Based on Quantized Hall Resistance. *Physical Review Letters* **45**, 494-

|   |   |
|---|---|
| | 497 (1980). |
| 2 | Tsui, D. C., Stormer, H. L. & Gossard, A. C. Two-Dimensional Magnetotransport in the Extreme Quantum Limit. *Physical Review Letters* **48**, 1559-1562 (1982). |
| 3 | Novoselov, K. S. *et al.* Two-dimensional gas of massless Dirac fermions in graphene. *Nature* **438**, 197 (2005). |
| 4 | Zhang, Y., Tan, Y.-W., Stormer, H. L. & Kim, P. Experimental observation of the quantum Hall effect and Berry's phase in graphene. *Nature* **438**, 201 (2005). |
| 5 | Bertrand, I. H. Possible States for a Three-Dimensional Electron Gas in a Strong Magnetic Field. *Japanese Journal of Applied Physics* **26**, 1913 (1987). |
| 6 | Kohmoto, M., Halperin, B. I. & Wu, Y.-S. Diophantine equation for the three-dimensional quantum Hall effect. *Physical Review B* **45**, 13488-13493 (1992). |
| 7 | Montambaux, G. & Littlewood, P. B. "Fractional" Quantized Hall Effect in a Quasi-One-Dimensional Conductor. *Physical Review Letters* **62**, 953-956 (1989). |
| 8 | Bernevig, B. A., Hughes, T. L., Raghu, S. & Arovas, D. P. Theory of the Three-Dimensional Quantum Hall Effect in Graphite. *Physical Review Letters* **99**, 146804 (2007). |
| 9 | Druist, D. P., Turley, P. J., Maranowski, K. D., Gwinn, E. G. & Gossard, A. C. Observation of Chiral Surface States in the Integer Quantum Hall Effect. *Physical Review Letters* **80**, 365-368 (1998). |
| 10 | Störmer, H. L., Eisenstein, J. P., Gossard, A. C., Wiegmann, W. & Baldwin, K. Quantization of the Hall effect in an anisotropic three-dimensional electronic system. *Physical Review Letters* **56**, 85-88 (1986). |
| 11 | Balents, L. & Fisher, M. P. A. Chiral Surface States in the Bulk Quantum Hall Effect. *Physical Review Letters* **76**, 2782-2785 (1996). |
| 12 | MacDonald, A. H. & Bryant, G. W. Strong-magnetic-field states of the pure electron plasma. *Physical Review Letters* **58**, 515-518 (1987). |
| 13 | Bardasis, A. & Das Sarma, S. Peierls instability in degenerate semiconductors under strong external magnetic fields. *Physical Review B* **29**, 780-784 (1984). |
| 14 | Yoshioka, D. & Fukuyama, H. Electronic Phase Transition of Graphite in a Strong Magnetic Field. *Journal of the Physical Society of Japan* **50**, 725-726 (1981). |
| 15 | Celli, V. & Mermin, N. D. Ground State of an Electron Gas in a Magnetic Field. *Physical Review* **140**, A839-A853 (1965). |
| 16 | Jérome, D., Rice, T. M. & Kohn, W. Excitonic Insulator. *Physical Review* **158**, 462-475 (1967). |
| 17 | Hannahs, S. T., Brooks, J. S., Kang, W., Chiang, L. Y. & Chaikin, P. M. Quantum Hall effect in a bulk crystal. *Physical Review Letters* **63**, 1988-1991 (1989). |
| 18 | Cooper, J. R. K., W. Auban, P. Montambaux, G. Jérome, D. and Bechgaard,K. Quantized Hall effect and a new field-induced phase transition in the organic superconductor (TMTSF)$_2$(PF)$_6$. *Physical Review Letters* **63**, 1984-1987 (1989). |
| 19 | Hill, S. *et al.* Bulk quantum Hall effect in η-Mo$_4$O$_{11}$. *Physical Review B* **58**, 10778-10783 (1998). |
| 20 | Cao, H. *et al.* Quantized hall effect and Shubnikov-de Haas Oscillations in highly doped Bi$_2$Se$_3$: Evidence for layered transport of bulk carriers. *Physical Review Letters* **108**, 216803 (2012). |
| 21 | Masuda, H. *et al.* Quantum Hall effect in a bulk antiferromagnet EuMnBi$_2$ with magnetically confined two-dimensional Dirac fermions. *Science Advances* **2**, e1501117-e1501117 (2016). |
| 22 | Trier, F. *et al.* Quantization of Hall Resistance at the Metallic Interface between an Oxide Insulator and SrTiO$_3$. *Physical Review Letters* **117**, 096804 (2016). |
| 23 | Jones, T. E., Fuller, W. W., Wieting, T. J. & Levy, F. Thermoelectric power of HfTe$_5$ and ZrTe$_5$. *Solid* |

*State Communications* **42**, 793-798 (1982).

24 Weng, H., Dai, X. & Fang, Z. Transition-Metal Pentatelluride ZrTe5 and HfTe5: A Paradigm for Large-Gap Quantum Spin Hall Insulators. *Physical Review X* **4**, 011002 (2014).

25 Li, Q. *et al.* Chiral magnetic effect in $ZrTe_5$. *Nature Physics* **12**, 550-555 (2016).

26 Liang, T. *et al.* Anomalous Hall effect in ZrTe5. *Nature Physics* **14**, 451-455 (2018).

27 Yuan, X. *et al.* Observation of quasi-two-dimensional Dirac fermions in $ZrTe_5$. *Npg Asia Materials* **8**, e325 (2016).

28 Yu, W. *et al.* Quantum Oscillations at Integer and Fractional Landau Level Indices in Single-Crystalline ZrTe5. *Scientific Reports* **6**, 35357 (2016).

29 Wu, R. *et al.* Evidence for Topological Edge States in a Large Energy Gap near the Step Edges on the Surface of ZrTe5. *Physical Review X* **6**, 021017 (2016).

30 Pariari, A. & Mandal, P. Coexistence of topological Dirac fermions on the surface and three-dimensional Dirac cone state in the bulk of ZrTe5 single crystal. *Scientific Reports* **7**, 40327 (2017).

31 Zhang, Y. *et al.* Electronic evidence of temperature-induced Lifshitz transition and topological nature in $ZrTe_5$. *Nature Communications* **8**, 15512 (2017).

32 Liu, Y. *et al.* Zeeman splitting and dynamical mass generation in Dirac semimetal $ZrTe_5$. *Nature Communications* **7**, 12516-12516 (2016).

33 Zheng, G. *et al.* Transport evidence for the three-dimensional Dirac semimetal phase in $ZrTe_5$. *Physical Review B* **93**, 115414 (2016).

34 Chen, R. Y. *et al.* Magnetoinfrared Spectroscopy of Landau Levels and Zeeman Splitting of Three-Dimensional Massless Dirac Fermions in $ZrTe_5$. *Physical Review Letters* **115**, 176404 (2015).

35 Wilson, J. A., Di Salvo, F. J. & Mahajan, S. Charge-density waves and superlattices in the metallic layered transition metal dichalcogenides. *Advances in Physics* **24**, 117 (1975).

36 McMillan, W. L. Theory of discommensurations and the commensurate-incommensurate charge-density-wave phase transition. *Physical Review B* **14**, 1496 (1976).

37 Lee, P. A. & Rice, T. M. Electric field depinning of charge density waves. *Physical Review B* **19**, 3970 (1979).

38 Monçeau, P., Ong, N. P., Portis, A. M., Meerschaut, A. & Rouxel, J. Electric Field Breakdown of Charge-Density-Wave---Induced Anomalies in $NbSe_3$. *Physical Review Letters* **37**, 602 (1976).

39 Andrei, E. Y. *et al.* Observation of a Magnetically Induced Wigner Solid. *Physical Review Letters* **60**, 2765-2768 (1988).

40 Zhang, C., Du, R.-R., Manfra, M. J., Pfeiffer, L. N. & West, K. W. Transport of a sliding Wigner crystal in the four flux composite fermion regime. *Physical Review B* **92**, 075434 (2015).

41 Yang, K., Haldane, F. D. M. & Rezayi, E. H. Wigner crystals in the lowest Landau level at low-filling factors. *Physical Review B* **64**, 081301 (2001).

42 Field, S. B., Reich, D. H., Rosenbaum, T. F., Littlewood, P. B. & Nelson, D. A. Electron correlation and disorder in $Hg_{1-x}Cd_xTe$ in a magnetic field. *Physical Review B* **38**, 1856-1864 (1988).

43 Shayegan, M., Goldman, V. J. & Drew, H. D. Magnetic-field-induced localization in narrow-gap semiconductors $Hg_{1-x}Cd_xTe$ and InSb. *Physical Review B* **38**, 5585-5602 (1988).

44 Ghezzi, C., Magnanini, R. & Parisini, A. Metal–insulator transition induced by the magnetic field in n-type GaSb. *Solid State Communications* **136**, 126 (2005).

45 Hopkins, P. F., Burns, M. J., Rimberg, A. J. & Westervelt, R. M. Magnetic-field-induced

localization in degenerately doped n-type Ge. *Physical Review B* **39**, 12708 (1989).

## Acknowledgements

The authors thank Jiawei Mei, Liang Fu, Xi Dai, Xiangrong Wang, K.T. Law, Dapeng Yu, Fan Zhang, Jinqi Wu, and D. L. Deng for valuable discussions. This work was supported by Guangdong Innovative and Entrepreneurial Research Team Program (No.2016ZT06D348) and Shenzhen Fundamental subject research Program (JCYJ20170817110751776) and (JCYJ20170307105434022). Work at USTC was supported by the National Key R & D Program of China (2016YFA0301700 and 2017YFB0405703), NNSFC (No. 11474265), and China Government Youth 1000-Plan Talent Program. Work at Brookhaven is supported by the Office of Basic Energy Sciences, U.S. Department of Energy under Contract No. DE-SC0012704. KY's work is supported by DOE grant No. de-sc0002140, and National Science Foundation Cooperative Agreement No. DMR-1644779, and the State of Florida. PAL acknowledges support by the US Department of Energy Basic energy sciences under grant DE-FG02-03-ER46076. Work at Singapore was supported by the Singapore Ministry of Education AcRF Tier 2 (MOE2015-T2-2-144).

## Author contributions

F.D.T and Y.F.R equally contributed to this work. F.D.T and P.P.W carried out the transport measurements. J.S, R.D.Z, and G.G. prepared the samples. Y.F.R, S.A.Y, K.Y, and P.A.L. performed the theoretical analysis. Z.Q and L.Z supervised this work. All authors contributed to the interpretation and analysis of the data and the writing of the manuscript.

## Additional information

The authors declare no competing financial interests.

**Supplementary information** is available in the online version of the paper.

**Data availability.** The data that support the findings of this study are available from the corresponding author on request.

**Methods**

**Sample synthesis and characterizations.** High-quality single crystal ZrTe5 was synthesized with high-purity elements ( 99.9999% Zirconium and 99.9999% Tellurium), and the needle-like crystals ($\sim 0.1 \times 0.3 \times 20 \, \text{mm}^3$) were obtained by the Tellurium flux method and Chemical Vapor transport. The lattice parameters of Crystals were structurally confirmed by X-ray diffraction, scanning tunneling microscopy, and transmission electron microscopy with electron diffraction. The low-temperature magneto-transport measures were performed in an Oxford TeslatronPT cryostat with variable temperature range from 1.5 K to 300 K, and rotatable Swedish insert with angles between 0 and 290 degrees, or in a ³He insert refrigerator with the base temperature down to 260 mK, by a superconductor magnetic field up to 14 T. In all our measurements, the current $I$ (10 ~ 100 μA) was applied along an axis, and four-terminal resistance was measured using the standard lock-in method with a low frequency (17.777 Hz). The magnetic field is rotated from $z$ axis to both $x$ or $y$ axis to measure the anisotropy. The bulk carrier densities $n \sim 10^{16} \, \text{cm}^{-3}$ were coincidental obtained from the Hall effect measurements at the low magnetic field, and the Fermi packets from the analysis of SdH Oscillations in $x$, $y$ and $z$ axes.

**Table *I*:**

| Parameters | $x$ | $z$ | $y$ |
|---|---|---|---|
| Cyclotron mass $m_i^*/m_e$ | 0.344±0.008 | 0.016±0.004 | 0.12±0.01 |
| Effective mass $M_i/m_e$ | 0.006±0.002 | 2.5±0.9 | 0.05±0.02 |
| Oscillation Frequency $B_F$ (T) | 15.7±0.2 | 1.18±0.02 | 9.2±0.1 |
| Fermi area $S_F$ ($10^{-4}$ Å$^{-2}$) | 15.0±0.2 | 1.13±0.01 | 8.8±0.2 |
| $k_F$ ($10^{-3}$ Å$^{-1}$) | 4.6±0.3 | 61±4 | 7.8±0.6 |
| $v_F$ ($10^5$ m/s) | 9±3 | 0.3±0.1 | 1.9±0.7 |
| Lifetime $\tau$ (ps) | 0.8±0.2 | 0.41±0.03 | 0.38±0.02 |

**Table *I* | Charge transport parameters.** Anisotropic transport parameters along different directions.

**Table *II*:**

| Sample | S#1 | S#2 | S#3 | S#4 |
|---|---|---|---|---|
| $\lambda_{F,z}$ (nm) | 5.40±0.03 | 11.30±0.07 | 9.3±0.3 | 10.2±0.2 |
| $\lambda_Q$ (nm) | 3.1±0.5 | 5.9±0.6 | 5±1 | 5.2±0.5 |
| $\lambda_Q/\lambda_{F,z}$ | 0.57±0.09 | 0.52±0.05 | 0.5±0.1 | 0.51±0.06 |
| $\lambda_Q/(7.25$Å$)$ | 4.2±0.4 | 8.1±0.8 | 6.9±1.4 | 7.2±0.8 |

**Table *II* | Fermi wavelength $\lambda_{F,z}$ and charge density wave period $\lambda_Q$ for different samples.**

# Figure

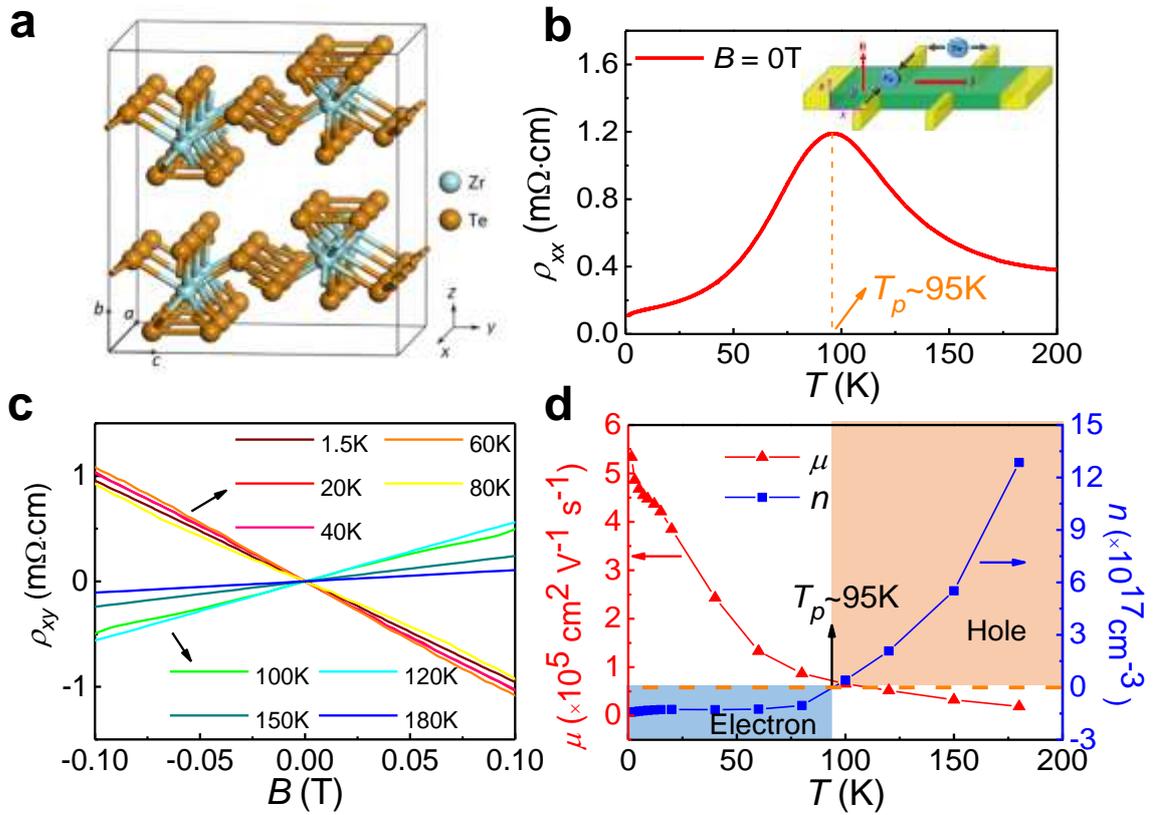

**Figure 1 | a**, The crystal structure of ZrTe$_5$. **b**, The temperature-dependent resistivity $\rho_{xx}(T)$ at zero magnetic fields. The anomalous resistance peak occurs around $T_p \sim 95$ K. The inset shows the sample with Hall bar contact. **c**, Temperature-dependent Hall resistivity of ZrTe$_5$. **d**, The temperature-dependent mobility $\mu$ and carrier density $n$ of a dominant carrier. A transition of electron- to hole-dominated transport is observed around the temperature $T_p$.

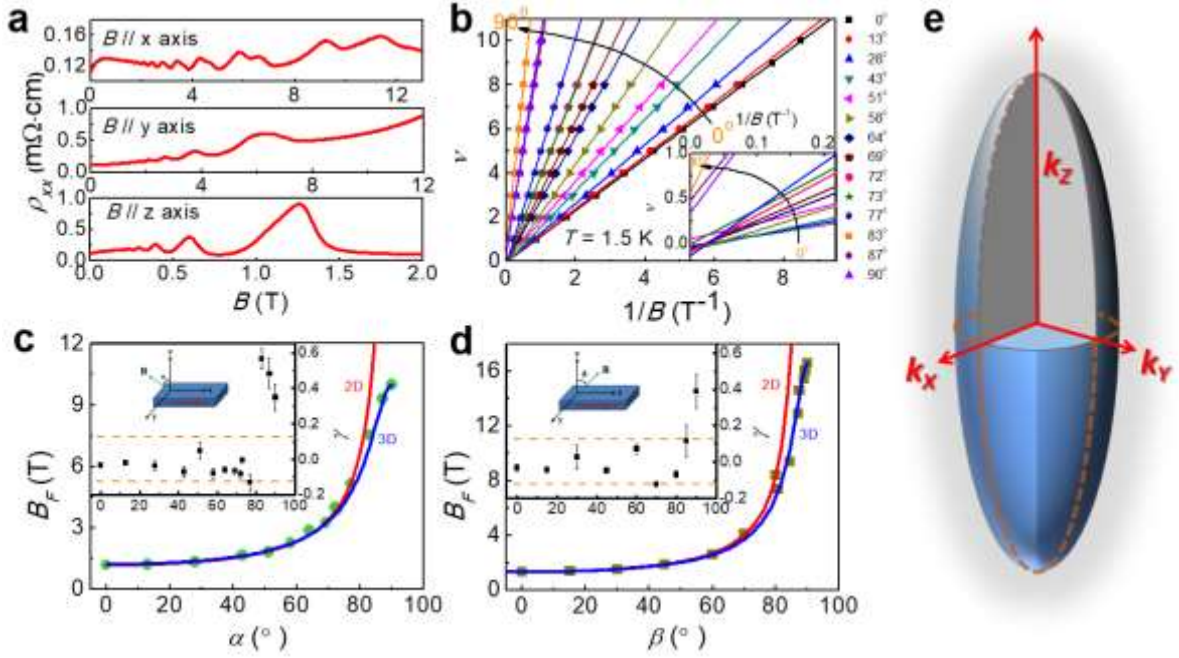

**Figure 2 | The Fermi surface's topology and morphology in ZrTe₅. a**, Quantum SdH oscillation with $B \parallel x, y$, and $z$ axis for longitudinal resistivity $\rho_{xx}(B)$. **b**, Landau fan diagram of LL index $N$ versus $1/B$ for different angles of $\beta$ that lies in $x$-$z$ plane and is tuned from $z$ to $x$-direction. **c**, **d**, Angle dependence of oscillation frequency $B_F$ vs angles θ (θ=α, or β), the red fitting curves represent the planar 2D Fermi surface, and blue fitting curves are for 3D ellipsoidal Fermi surface. Inset: the angle dependence of intercept of Landau fan diagram with error bar. **e**, The topology and morphology of ZrTe₅'s Fermi surface in momentum space with $k_{F,x}$, $k_{F,y}$, and $k_{F,z}$.

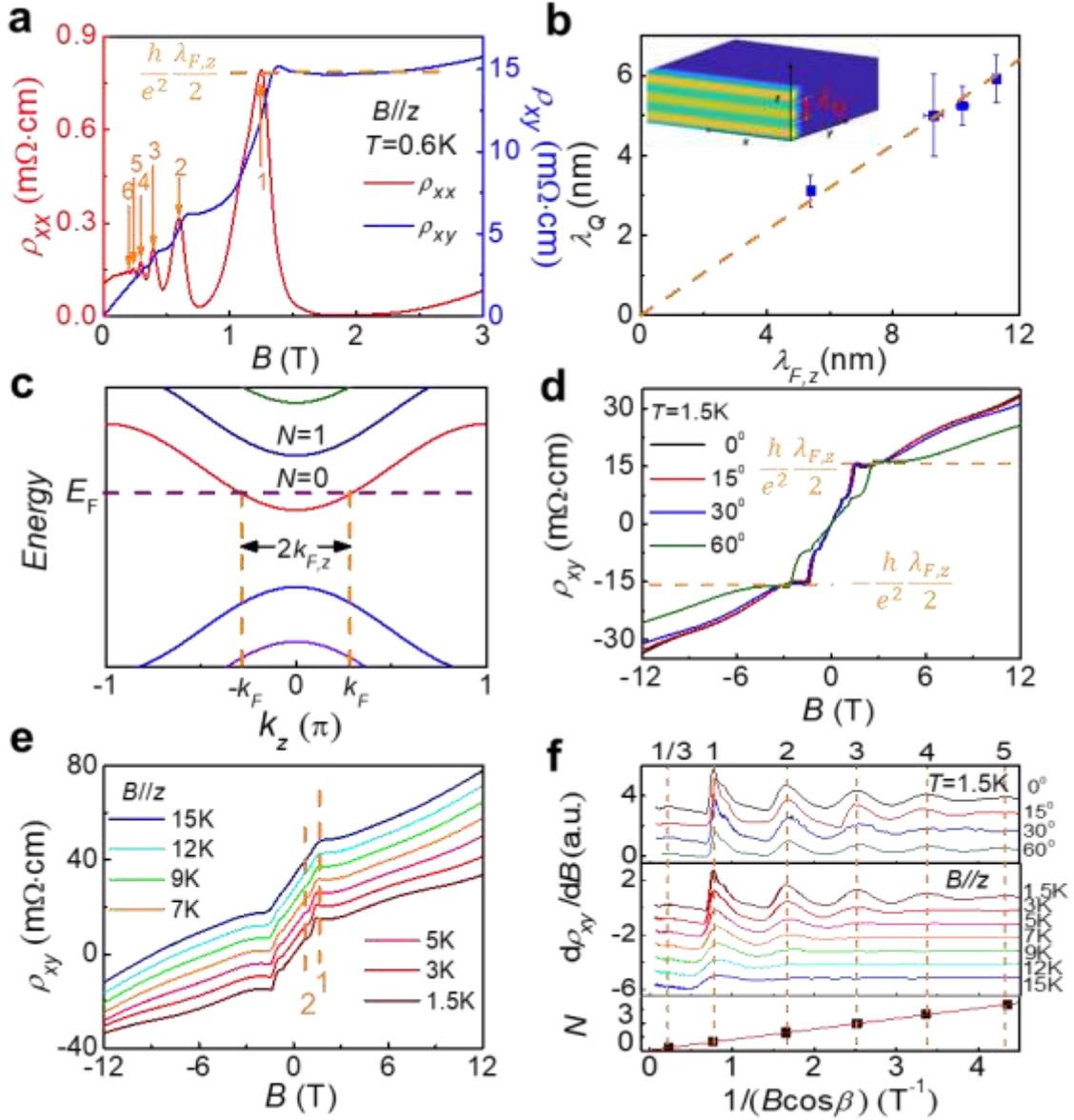

**Figure 3 | Three-dimensional Quantum Hall effect and transport signatures of edge states. a**, Longitudinal resistivity, $\rho_{xx}(B)$ (red, left scale), and Hall resistivity $\rho_{xy}(B)$ (blue, right scale) as a function of magnetic field at temperature 0.6 K. **b**, Periodic of CDW $\lambda_Q$ vs Fermi wavelength $\lambda_{F,z}$ in $z$-direction. The inset is an illustration of 2D conducting layer simulation. **c**, Schematical plot of Landau-level dispersion along $z$ direction and Fermi wavevector. **d**, Hall resistivity $\rho_{xy}(B)$ vs magnetic field $B$ applied along different directions $\alpha$ tuning from $z$ direction to $x$ direction. **e**, Hall resistivity $\rho_{xy}$ vs $B$ under different temperatures. Lines with neighbor temperature is shifted by 5 mΩ·cm for clarity. **f**, The $-d\rho_{xy}(B)/dB$ vs $1/B_\perp$. Up panel shows results for magnetic field along different angles of $\beta = 0°, 15°, 30°$ and $60°$ with $B_\perp = B\cos\beta$. Middle panel shows results at different temperatures the same as panel **e**. Vertical dash lines are drawn in **e** and **f** corresponding to the Landau level filling factor $\nu$. Bottom panel: Landau fan diagram of LL index $N$ vs $1/B_\perp$.

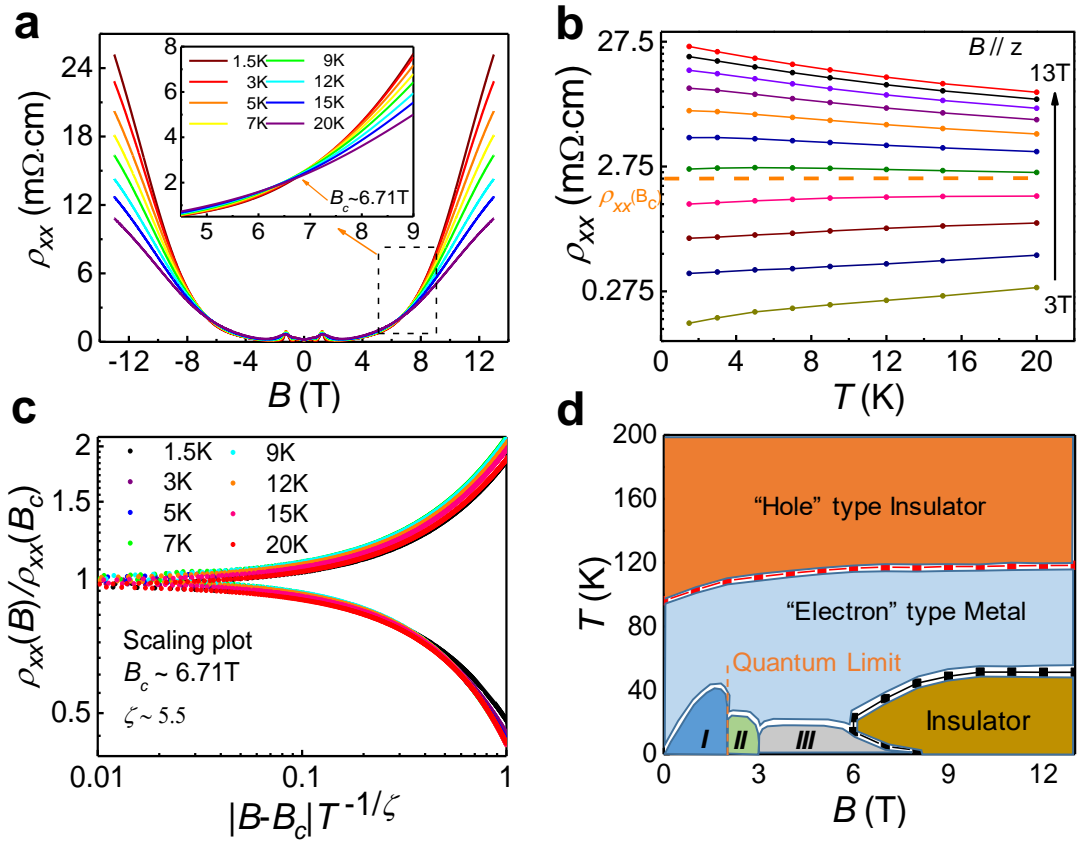

**Figure 4 | Metal-insulator transition and phase diagram with ($T$, $B$). a**, Longitudinal resistance $\rho_{xx}(B)$ vs magnetic field $B$ under different temperature $T$ ranging from 1.5 to 20 K. **b**, $\rho_{xx}(B)$ vs temperature $T$ with magnetic field $B$ varying from 3 to 13 T. **c**, Normalized resistance $\rho_{xx}(B)/\rho_{xx}(B_c)$ vs scaled parameter $|B - B_c|T^{-1/\zeta}$. **d**, Schematic plot of phase diagram summarizing our results. See main text for discussions.